\documentclass[%
 reprint,
 floatfix,
superscriptaddress,
longbibliography,
 amsmath,amssymb,
 aps,
]{revtex4-2}

\makeatletter
\def\set@curr@file#1{%
 \begingroup
 \escapechar\m@ne
 \xdef\@curr@file{\expandafter\string\csname #1\endcsname}%
 \endgroup
}
\def\quote@name#1{"\quote@@name#1\@gobble""}
\def\quote@@name#1"{#1\quote@@name}
\def\unquote@name#1{\quote@@name#1\@gobble"}
\makeatother

\usepackage{graphicx}% Include figure files
\graphicspath{ {./images/} }
\usepackage{dcolumn}% Align table columns on decimal point
\usepackage{bm}% bold math
\usepackage{xfrac}
\usepackage{siunitx}
\usepackage{amsmath}
\usepackage[utf8]{inputenc}
\usepackage[T1]{fontenc}
\usepackage[export]{adjustbox}
\usepackage[colorinlistoftodos,prependcaption,textsize=tiny]{todonotes}
\usepackage{gensymb}

\usepackage{upgreek}

\presetkeys%
 {todonotes}%
 {inline, backgroundcolor=yellow}{}
\begin{document}

\preprint{APS/123-QED}

\title{Evaluation of gas permeability in porous separators for polymer electrolyte fuel cells: CFD simulation based on micro X-ray computed tomography images}% Force line breaks with \\

\author{Soichiro Shimotori}
 \email{soichiro.shimotori@toshiba.co.jp}
 \altaffiliation[Also at ]{Department of Mechanical Engineering, The University of Tokyo.}%Lines break automatically or can be forced with \\
 \affiliation{Toshiba Energy Systems $\&$ Solutions Corporation, 4-1 Ukishima-cho, Kawasaki-ku, Kawasaki 210-0862, Japan
}%

\author{Toshihiro Kaneko}
\author{Yuta Yoshimoto}
\author{Ikuya Kinefuchi}
\author{Amer Alizadeh}
\author{Wei-Lun Hsu}
 \affiliation{Department of Mechanical Engineering, The University of Tokyo, 7-3-1 Hongo, Bunkyo-ku, Tokyo 113-8656, Japan
}%

\author{Hirofumi Daiguji}
 \email{Corresponding author: daiguji@thml.t.u-tokyo.ac.jp}
 \affiliation{Department of Mechanical Engineering, The University of Tokyo, 7-3-1 Hongo, Bunkyo-ku, Tokyo 113-8656, Japan
}%

\date{\today}% It is always \today, today,

\begin{abstract}
Pore structures and gas transport properties in porous separators for polymer electrolyte fuel cells  are evaluated both  experimentally and through simulations. In the experiments, the gas permeabilities of two porous samples, a conventional sample and one with low electrical resistivity, are measured by a capillary flow porometer, and the pore size distributions are evaluated with mercury porosimetry. Local pore structures are directly observed with micro X-ray computed tomography (CT). In the simulations, the effective diffusion coefficients of oxygen and the air permeability in porous samples are calculated using random walk Monte Carlo simulations and computational fluid dynamics (CFD) simulations, respectively, based on the X-ray CT images. The calculated porosities and air permeabilities of the porous samples are in good agreement with the experimental values. The simulation results also show that the in-plane permeability is twice the through-plane permeability in the conventional sample, whereas it is slightly higher in the low-resistivity sample. The results of this study show that CFD simulation based on micro X-ray CT images makes it possible to evaluate anisotropic gas permeabilities in anisotropic porous media.
\end{abstract}

\maketitle

%\tableofcontents

\section{\label{sec:level11}INTRODUCTION}

Fuel cells are promising electrochemical devices that use the reaction of hydrogen and oxygen to generate electrical power and are regarded as clean and low-cost power sources. These environmentally friendly chemical cells show a high energy conversion efficiency without the limitation imposed by  the Carnot cycle. Owing to their excellent electrical conductivity and durability, polymer electrolyte fuel cells (PEFCs), which utilize a solid polymer thin membrane as electrolyte between the electrodes to prevent the  leakage problems that occur in liquid electrolyte chemical cells, have attracted considerable attention over the past few years~\cite{springer1991,azweber2004}. In general, PEFCs consist of a polymer membrane, catalyst layers, microporous layers, gas diffusion layers (GDLs), gas channels (GCs) and separators. Owing to the wide operating temperature range from room temperature to 100$^\circ$C~\cite{bernardi1990}, PEFCs can be used for a variety of applications, from large-scale facilities to portable systems such as engines for the next generation of electric vehicles, in which they are able to achieve a minimum response time for  engine start-up and shutdown. 

Water produced in PEFCs from the cathode reaction between protons and oxygen has a remarkable effect on the ion conductivity of the polymer electrolyte membrane and thus the performance of the cell. The water content of the membrane is subject to the surrounding vapor pressure and the water transport in the cell. In addition to  water penetration from the cathode, water is electro-osmotically transported from  anode to cathode owing to the protons (positively charged) contained in the membrane~\cite{zawodzinski1995}. This gives rise to a dehydrated membrane, reducing the ion conductivity, and to an overhydrated cathode, hindering the oxygen supply to the cathode and thus resulting in a deterioration in cell efficiency. Therefore, to maximize the ion conductivity and performance of PEFCs,  water management is an important task, but one that has yet to be optimized. 

\begin{figure}[t!]
\includegraphics[width=\columnwidth]{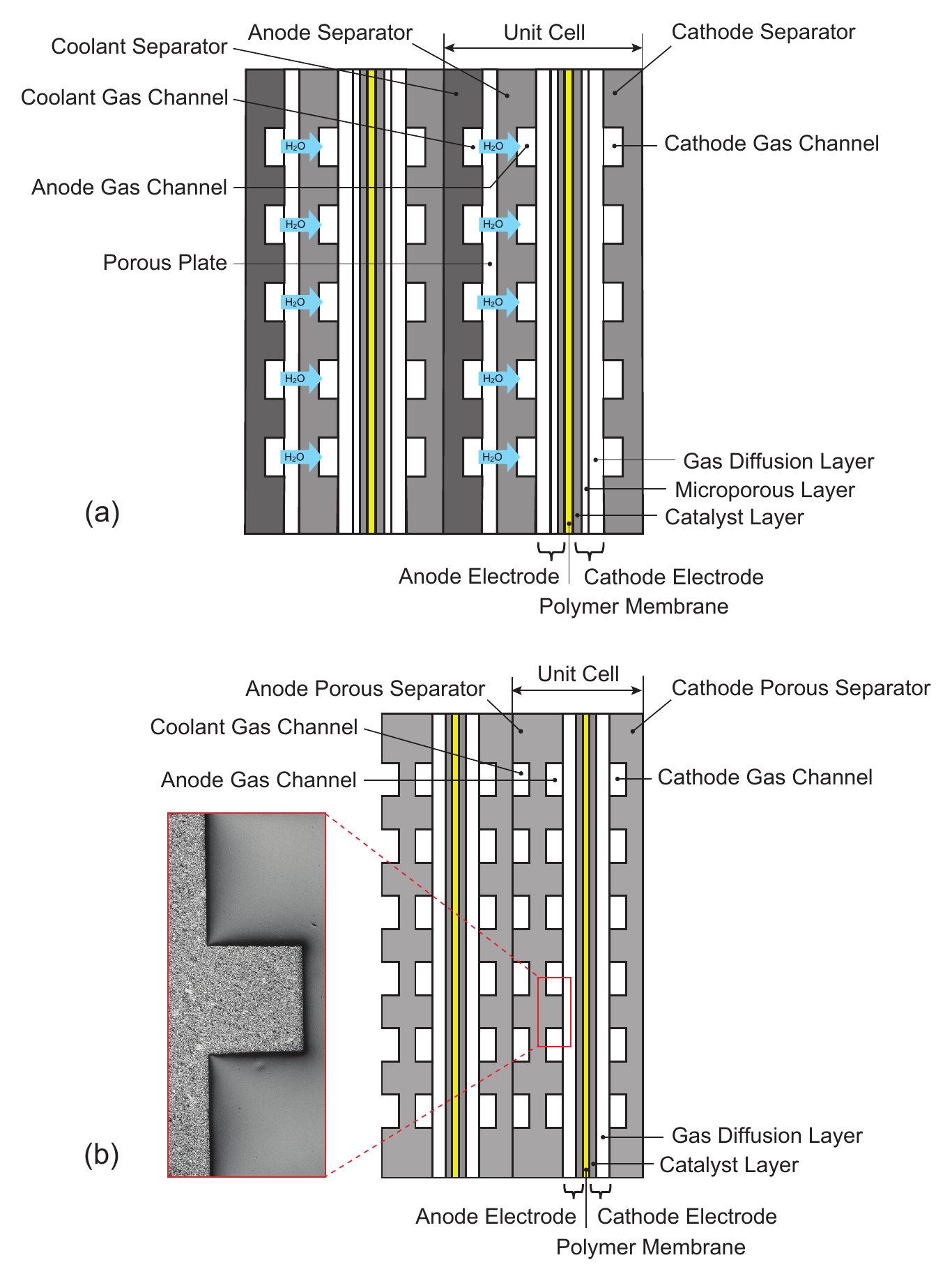}%
\caption{\label{fig:schem1} (a) Schematics of (a) an internally humidified PEFC and (b) a PEFC with porous separators. The inset shows a scanning electron microscope image of a porous separator.}
\end{figure}

Humidification of the incoming reactant gases (external humidification) is one method to control the water content of PEFCs. To compensate for electro-osmotically transported water, humidification designs for the anode stream have been proposed~\cite{nguyen1993}. Humidification of reactant gases at a temperature higher than the cell temperature has also been recommended~\cite{ticianelli1988}. In addition, the water balance across the cell and its sensitivity to changes in operating conditions have been studied~\cite{bernardi1990, wilkinson1994, fuller1993}. Self-humidification is another method to control the water content of PEFCs. This method employs a total heat and mass exchanger between  the supply oxidant and the exhaust oxidant~\cite{shimotori1999}, which transfers  vapor from  exhaust  to  supply oxidant without any extra energy consumption. Yet another method for humidification is to  supply water directly to the cell (internal humidification) by inserting a porous plate, which allows the supply of water from the coolant channels to be transferred to the anode without extra energy consumption~\cite{wood1998, shimotori1996}, as illustrated in Fig.~\ref{fig:schem1}(a).

As already mentioned, mitigation of the flooding phenomenon in an overhydrated cathode is also important to prevent deterioration in cell efficiency. One method is to introduce porous separators at  both cathode and anode, as illustrated in Fig.~\ref{fig:schem1}(b). As long as the pressure in the coolant channel is lower than that in the reactant gas channel, water can be continuously removed from the cathode through GDLs~\cite{baldwin1990, zaffou2006, yi2004}. The incorporation of both an internal humidification unit and porous separators gives enhanced performance compared with conventional PEFCs using only  external humidification, by providing a high water content in the polymer electrolyte membrane while  removing a sufficient amount of liquid water from the cathode~\cite{weber2007}.

%\begin{figure}[!t]
%\includegraphics[width=\columnwidth]{fig_2}%
%\caption{\label{fig:schem2} Schematic  of a PEFC with porous separators. The inset shows a scanning electron microscope image of a porous separator.}
%\end{figure}

In this context, the characteristics of water transfer across the porous separators becomes a key factor in the control of water supply and removal in PEFCs.  It is important to enhance the water permeability while simultaneously maintaining a competitive electrical conductivity. However, the correlation between  pore structure on the one hand and  water permeability and electrical resistivity on the other has yet to be clarified. In general, the water permeability through a porous medium is defined by the product of the  relative permeability of the liquid and the absolute permeability, where the  relative permeability  is a function of water saturation while the absolute permeability is an intrinsic property determined by  the pore structure alone~\cite{weber2004, bear1988}. Therefore, to characterize  water transport through a porous separator, evaluation of the absolute permeability of the porous medium is essential. Recently, the characterization of porous structures in PEFCs from the micro- to the nanoscale has been carried out  using advanced tomographic techniques. For instance, catalyst layers have a pore structure of only a few tens of nanometers. Several studies have carried out three-dimensional (3D) transmission electron microscope (TEM) tomography to visualize the nanoscale structure~\cite{okumura2017}. Microporous layers have sub-micrometer-scale and micrometer-scale pore structures. 3D scanning electron microscope (SEM) tomography has  commonly been used to visualize the sub-micrometer-scale structure~\cite{kanno2011, iwai2010}. On a larger scale of micrometers, 3D X-ray computed tomography (CT) has been employed~\cite{kinefuchi2013, grew2010}.

In the present study, 3D X-ray CT is performed to investigate the micrometer-scale pore structure of porous separators with different electrical resistivities. The scanned images of a 3D cylinder with diameter 1~mm   and height 1~mm  are binarized and the porosities of porous samples are evaluated. Then, the 3D cylindrical image is divided into cubes of side 200~$\upmu$m and the porosity of each cube is calculated. After porosity uniformity has been confirmed, the structural data for the central cube are transferred to voxel data for further calculations of transport properties. The pore size distribution for the central cube is confirmed to be the same to that measured with mercury porosimetry, and then effective diffusion coefficients of oxygen and air permeability are calculated with random walk Monte Carlo simulations and computational fluid dynamics (CFD) simulations, respectively. The objective of the former simulations is to evaluate the structural properties of porous separators such as anisotropy and tortuosity, while that of the latter simulations is to evaluate the anisotropic air transport properties in porous separators. The flow rates and permeabilities of air through porous samples calculated by CFD are compared with those measured by a capillary flow porometer. Under real operating conditions of PEFCs, water will be in both  vapor and liquid states in the porous separator. The present study focuses on gas transport properties in a porous separator and clarifies the relationship between these properties and the porous structure. Analysis of the complex vapor--liquid two-phase flow in a porous separator is outside the scope of this study.

\section{\label{sec:level12}EXPERIMENTAL METHODS}

\subsection{\label{sec:level22a}Sample preparation}

A porous separator for PEFCs was made from carbon. Such separators are  usually made by pouring a mixture of carbon particles and resin into a patterned mold for a plate with small gas channels as shown in Fig.~\ref{fig:schem1}(b). However, in a patterned mold, the porous structure at the corners of the channels may  differ from that on the flat surface of the channels, because the local stress acting on the mixture of carbon particles and resin depends on the local geometry. Thus, for the following experiments, a flat mold was used for the fabrication of a flat porous separator to ensure the uniformity of the porous structure.

Two types of flat porous separators with different electrical resistivities, a conventional type (sample A) and a low-resistivity type (sample B), were fabricated. To reduce electrical resistivity between carbon particles, carbon nanoparticles were added to sample B. In the fabrication process of a flat porous separator, the mixture of carbon particles and resin (plus carbon nanoparticles in sample B) was pored into a flat mold and was pressed in a certain direction. For the following experiments, the press direction is called through-plane direction and the normal to the press direction is called in-plane direction. Each of these two samples was cut into small pieces with three different dimensions for three different experiments: (a) flow porometry, (b) mercury porosimetry, and (c) X-ray CT. The electrical resistivity of each sample and the dimensions of the flat porous separators and small pieces for the three different experiments are shown in Table~\ref{tab:table1}. Note that the thickness direction ($T$) is the through-plane direction, that is, the press direction and the length ($L$), width ($W$) and diameter ($D$) directions are the in-plane directions.

\begin{table}[!t]
\caption{\label{tab:table1}%
Electrical resistivities and dimensions of porous samples.}
\begin{ruledtabular}
\begin{tabular}{lcc}
& Sample A & Sample B\\
\hline
Electrical resistivity (m$\Omega$\,m) & 0.356 & 0.249\\
Dimensions (mm):&&\\
\quad Flat porous separator & \multicolumn{2}{c}{$320\,(L) \times 270\,(W) \times 2.5\,(T)$}\\
\quad (a) Flow porometry  & \multicolumn{2}{c}{$30\,(D) \times 2.5\,(T)$}\\
\quad (b) Mercury porosimetry  & \multicolumn{2}{c}{$50\,(L) \times 50\,(W) \times 2.5\,(T)$}\\
\quad (c) X-ray CT & \multicolumn{2}{c}{$10\,(L) \times 2\,(W) \times 2.5\,(T)$}\\
\end{tabular}
\end{ruledtabular}
\end{table}

\subsection{\label{sec:level22b}Permeability measurement}

The permeability of air through a porous sample in the thickness direction was measured by a capillary flow porometer (iPORE, Porous Materials Inc.)~\cite{jena2001}. The disk-shaped sample was placed in a cylinder test rig, and air was supplied to the sample within the pressure range from 0 to 900~kPa at room temperature. The air flow rate was measured by a mass flow meter in the air supply line. The permeability of air, $k$, can be calculated from the Darcy's law for gas flow with pressure-dependent gas density~\cite{dullien1992}:
\begin{equation}
 k = 
 \frac{F}{A} \mu \frac{T}{\Delta P\left( 1 + \Delta P/2P \right)},
\label{eq:one}
\end{equation}
where $F$ is the volumatic flow rate of air at the outlet, $A$ is the surface area of the sample, $\mu$ is the viscosity of air, $T$ is the thickness of the sample, $P$ is the outlet pressure which is kept at atomspheric pressure, and $\Delta P$ is the pressure difference between the inlet and the outlet.

\subsection{\label{sec:level22c}Pore size distribution measurement}

The pore size distribution of the porous samples was measured by a mercury porosimeter (AutoPore IV 9510, Micromeritics Inc.). The measured pore diameter ranged from $d_{p} = 0.003$ to 100~$\upmu$m. The contact angle and surface tension of mercury used in the calculation of pore diameter are 130$^\circ$ and 485~dyn/cm, respectively.

%\begin{table}[!t]
%\caption{\label{tab:table2}%
%Contact angle and surface tension of mercury.
%}
%\begin{ruledtabular}
%\begin{tabular}{ll}
%Contact angle & 130$^\circ$ \\
%Surface tension & 485~dyn/cm\\
%\end{tabular}
%\end{ruledtabular}
%\end{table}

\begin{figure}[!b]
\includegraphics[width=\columnwidth]{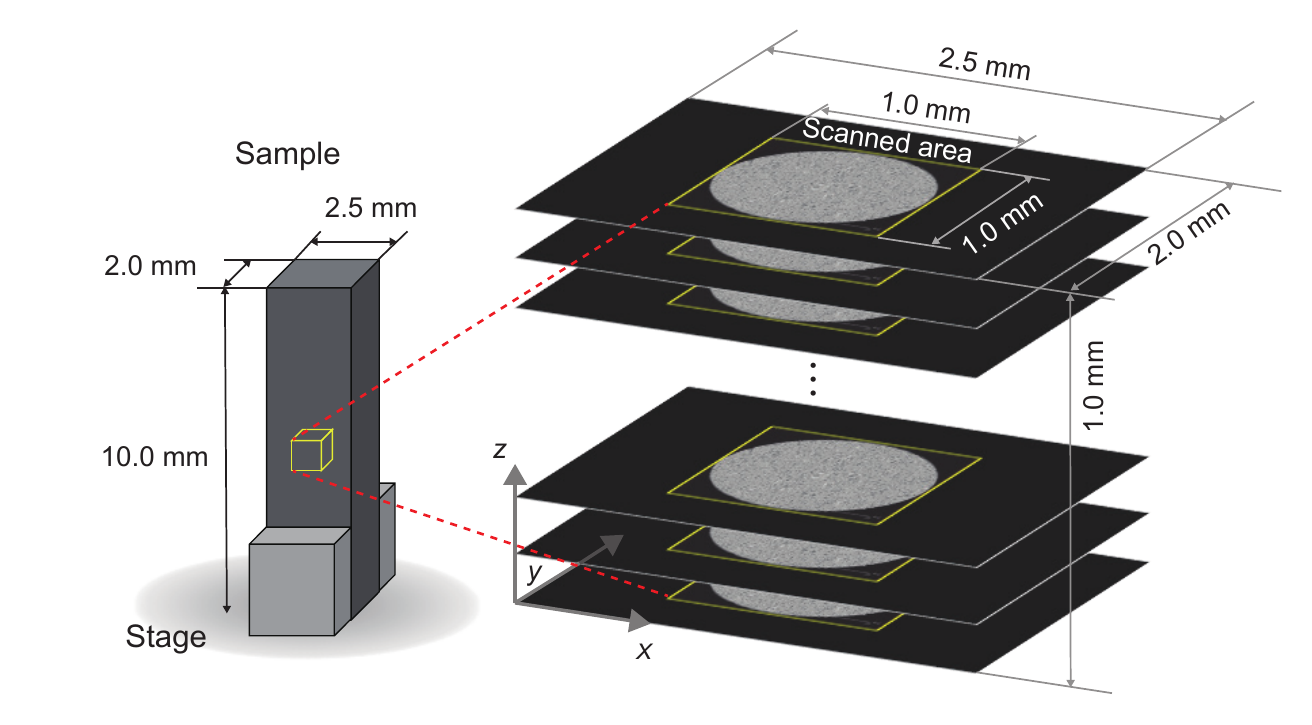}%
\caption{\label{fig:xct} Illustration of a porous sample on the stage and the cylindrical image scanned by X-ray CT.}
\end{figure}
\begin{figure}[!b]
\includegraphics[width=\columnwidth]{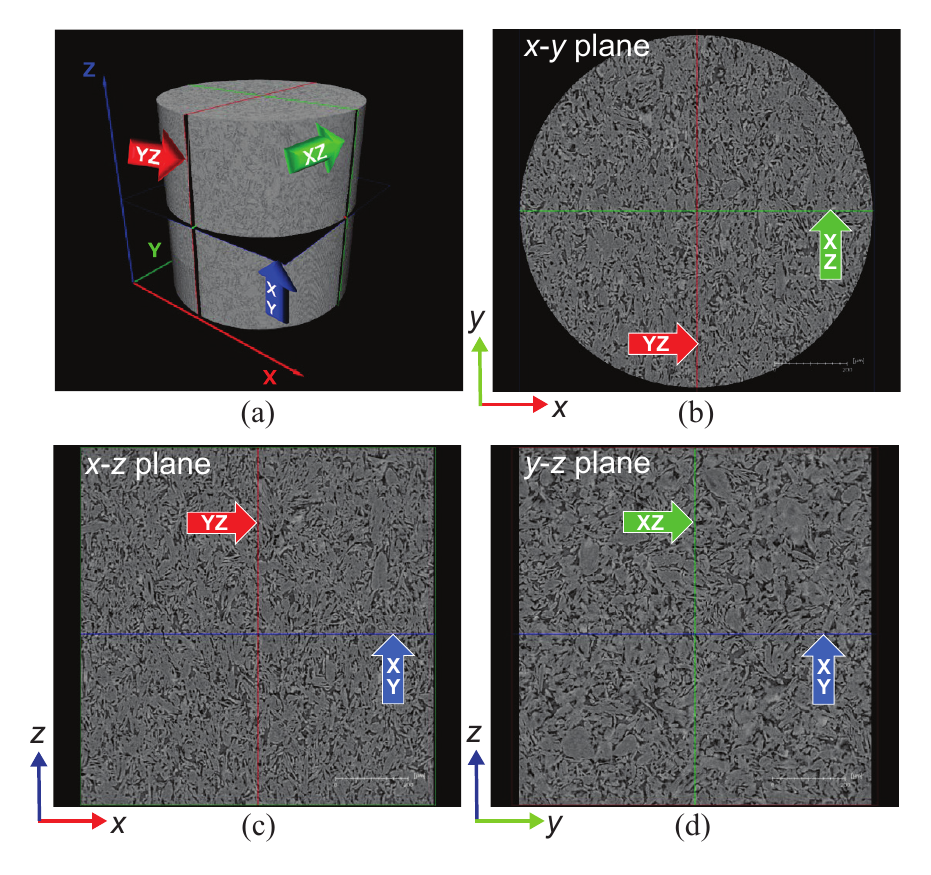}%
\caption{\label{fig:3d} (a) 3D image of a porous sample and its cross-sectional images in  (b) $x$--$y$, (c) $x$--$z$, and (d) $y$--$z$ planes (sample A).}
\end{figure}
%\begin{figure}[!t]
%\includegraphics[width=\columnwidth]{fig_4}%
%\caption{\label{fig:coord} Cross-sectional image of a scanned porous sample in the $x$--$y$ %plane.}
%\end{figure}

The 3D pore structure of a porous sample was reproduced using micro X-ray CT and image processing. The rectangular sample was set on the stage of a micro X-ray CT scanner (ZEISS Xradia 520 Versa, Carl Zeiss Inc.),  and 3D images were recorded by rotating the stage, as shown in Fig.~\ref{fig:xct}. An X-ray source was operated at 5~W (60~kV and 0.83~mA). The spatial resolution was 1~$\upmu$m per pixel for a 1~mm $\times$ 1~mm $\times$ 1~mm field of view. The central portion of the sample with dimensions of 10~mm~($L$) $\times$ 2~mm~($W$) $\times$ 2.5~mm~($T$)  was scanned, and the scanned data were used for image processing. Here, the $x$, $y$, and $z$  directions correspond to the thickness $T$, width $W$, and length $L$ of the sample, respectively. The scanned area in the $x$--$y$ plane was 1.0~mm $\times$ 1.0~mm. 995 cross-sectional images in the $x$--$y$ planes were stacked vertically in the $z$ direction. A 3D image of the sample in  Cartesian  coordinates, and the cross-sectional images in the $x$--$y$, $x$--$z$, and $y$--$z$ planes, are shown in Fig.~\ref{fig:3d}. It should be noted that the $y$--$z$ plane corresponds to the in-plane of the sample (the $x$ direction corresponds to the thickness $T$ of the sample), while the $x$--$z$ and $x$--$y$ planes correspond to the through-plane of the sample (the $y$ and $z$ directions correspond to the width $W$ and length $L$ of the  sample, respectively). In the scanned area, the high-brightness regions are the solid material consisting of carbon particles connected with resin, while the low-brightness regions are the pore regions. The remaining regions outside  the circle provide no information. The pore structure data of a 3D cylindrical region thus obtained were  used in the subsequent image processing

\subsection{\label{sec:level22d}X-ray computed tomography}

\begin{figure}[!b]
\includegraphics[width=\columnwidth]{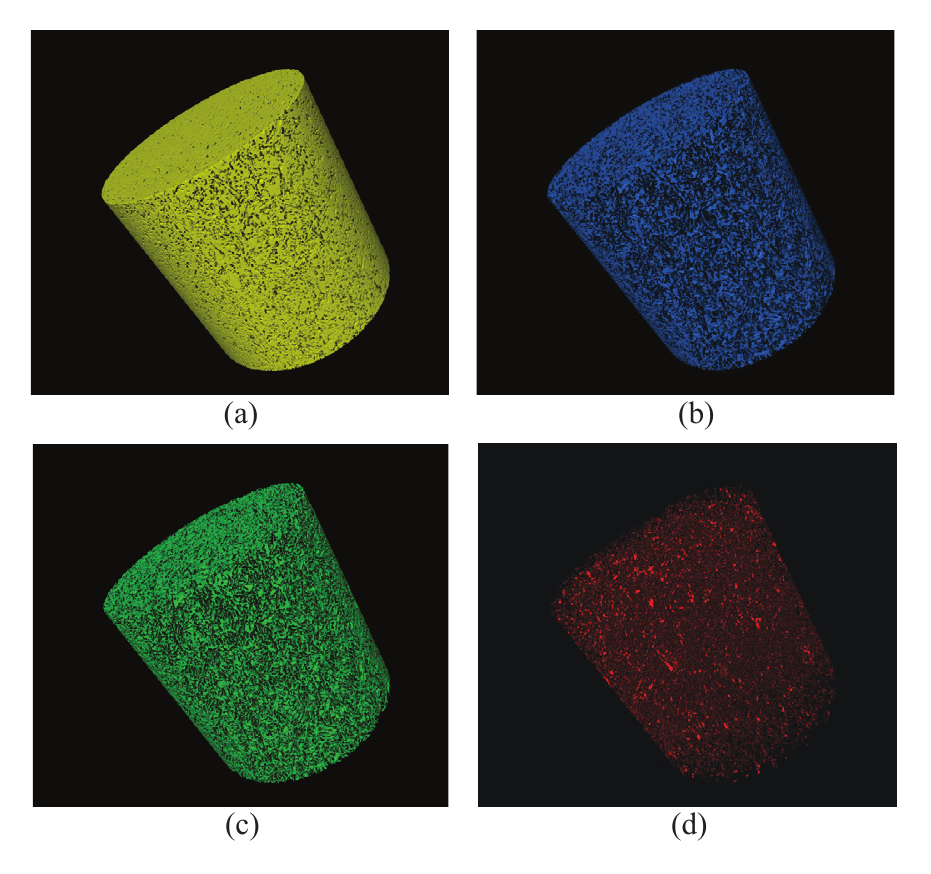}%
\caption{\label{fig:pore} Images of the entire cylindrical scanned region with 1~mm in diameter and 1~mm in height: (a) solid; (b) void; (c)  open pore; (d) closed pores (sample A).}
\end{figure}

In the image processing, the images were binarized by using \textquoteleft default\textquoteright~threshold in an image processing software ImageJ without considering any side information. The low-brightness regions below the threshold value were defined as pore regions, while the high-brightness regions above the threshold value were defined as solid regions. The number of pores and the volume of each pore were determined from the binarized image. The detected number of pores was 214\,514. The pore with the largest volume was taken as an open pore, with the remaining pores being closed pores. Images of solid, void, open pore, and  closed pores for the entire scanned area are shown in Fig.~\ref{fig:pore}.  The volumes of solid, void, open pore, and closed pores were $5.24\times10^{8}$, $2.02\times10^{8}$, $1.97\times10^{8}$, and $5.38\times10^{6}~\upmu$m$^{3}$, respectively. The porosity $\epsilon$ is given by
\begin{equation}
 \epsilon = 
  \frac{v_p}{v_p + v_s},
\label{eq:sixteen}
\end{equation}
where $v_p$  and $v_s$ are the pore and solid volumes, respectively.

%\begin{figure}[!t]
%\includegraphics[width=\columnwidth]{fig_6}%
%\caption{\label{fig:xy} (a) Cross-sectional image of a porous sample and (b) its binarized image %in the $x$--$y$ plane. The threshold brightness is 22\,500, the detected number of pores is 214\,514, and the volume of the largest pore is $1.97314\times10^{8}~\upmu$m$^3$.}
%\end{figure}

\section{\label{sec:level13}CALCULATION AND SIMULATION METHODS}

\subsection{\label{sec:level23a}Porosity distribution}

\begin{figure}[!b]
\includegraphics[width=\columnwidth]{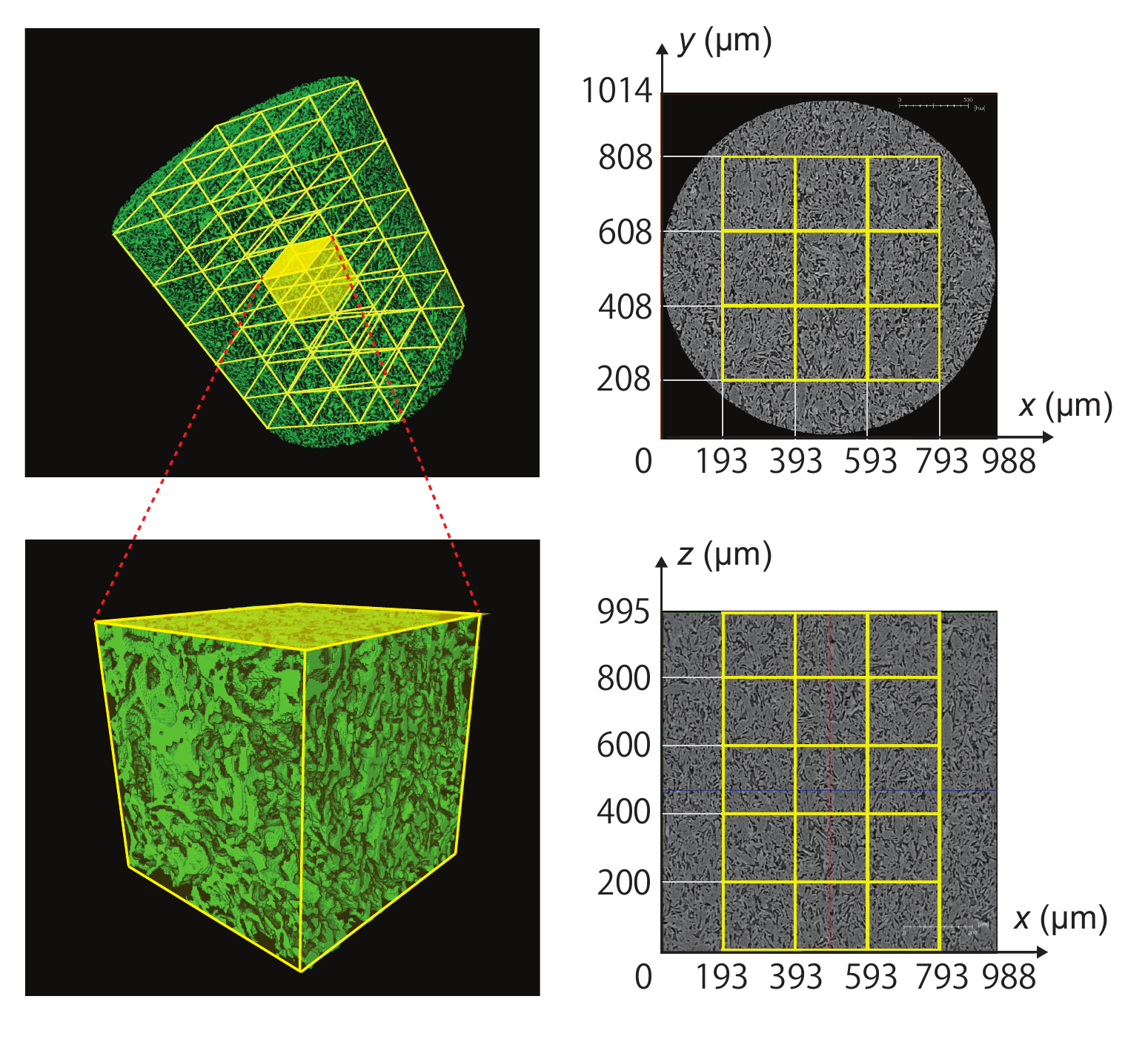}%
\caption{\label{fig:div} Cubic analysis region of $200\times200\times200$ voxels extracted from a cylindrical scanned region with 1~mm in diameter and 1~mm in height.}
\end{figure}

The following calculation and simulation were carried out  using part of the binarized image, a cubic region of $200\times200\times200$ voxels extracted from a cylindrical scanned region as shown in Fig.~\ref{fig:div}. A square area of $600\times600$ pixels within the circular scanned area was divided into nine square areas of $200\times200$ pixels. In the $z$ direction, the scanned area was divided into five areas. A total of 45 ($=9\times5$) cubic regions of $200\times200\times200$ voxels were thereby created in the entire scanned region as shown in Fig.~\ref{fig:div}. To evaluate the uniformity of the pore structure, the porosities of these 45 cubic regions were compared with each other. Each image was binarized at the same threshold level as mentioned in Sec.~\ref{sec:level22d} and a matrix  of brightness data [either 1 (solid) or 0 (pore)] was created. The number of solid pixels was counted in each square area and then integrated over the 200 images in the $z$ direction

\subsection{\label{sec:level23b}Effective oxygen diffusion coefficient}

In order to evaluate the anisotropy and tortuosity of porous samples, the MSD calculation was performed in the Kn diffusion region where the trajectory of an oxygen molecule is easily obtained without considering collisions between oxygen molecules. The effective diffusion coefficient for a porous structure, $D_e$, is given by~\cite{epstein1989, zalc2004, berson2011}
\begin{equation}
 D_e = 
  \frac{\epsilon D_0}{\tau},
\label{eq:two}
\end{equation}
where $\epsilon$ is the porosity, $\tau$ is the tortuosity factor, which represents the tortuous nature of a porous medium, and $D_0$ is a reference diffusion coefficient, typically described by the Bosanquet equation~\cite{pollard1948}:
\begin{equation}
 D_0 = 
  \left( \frac{1}{D_b} + \frac{1}{D_K} \right)^{-1},
\label{eq:three}
\end{equation}
where $D_b$ and $D_K$ are the diffusion coefficients in the bulk diffusion regime and the Knudsen diffusion regime, respectively. $D_b$ and $D_K$ are defined  using the mean free path $\lambda$, the characteristic length $d$, and the mean molecular velocity $v$ as follows:
\begin{align}
 D_b &= 
  \tfrac{1}{3} \lambda v,
\label{eq:four}
\\[3pt]
 D_K &= 
  \tfrac{1}{3} d v.
\label{eq:five}
\end{align}
By introducing the Knudsen number Kn, the following relation is satisfied:
\begin{align}
 D_b &= 
  \mathrm{Kn}\, D_K,
\label{eq:six}
%\\[3pt]
% D_K &= 
 % \mathrm{Kn}^{-1} D_b,
%\label{eq:seven}
\end{align}
where
\begin{equation}
 \mathrm{Kn} = 
  \frac{\lambda}{d}.
\label{eq:eight}
\end{equation}
When  $\mathrm{Kn} \rightarrow \infty$, the Bosanquet equation~(\ref{eq:three}) becomes
$D_0 \approx D_K$, and when $\mathrm{Kn} \rightarrow 0$, it becomes $D_0 \approx D_b$.
%\begin{equation}
% D_0 = 
%  \left( \frac{1}{\mathrm{Kn} \,D_K} + \frac{1}{D_K} \right)^{-1} = \frac{D_K}{1 + \mathrm{Kn}^{-1}} \approx D_K,
%\label{eq:nine}
%\end{equation}
%and when $\mathrm{Kn} \rightarrow 0$, it becomes
%\begin{equation}
% D_0 = 
%  \left( \frac{1}{D_b} + \frac{1}{\mathrm{Kn}^{-1}D_b} \right)^{-1} = \frac{D_b}{1 + \mathrm{Kn}} \approx D_b.
%\label{eq:ten}
%\end{equation}
The characteristic length $d$ was calculated using the following equation~\cite{weber2004,bear1988}:
\begin{equation}
 d =
  \left( \frac{\langle l^2 \rangle}{2 {\langle l \rangle}^2} - \beta \right) \langle l \rangle,
\label{eq:eleven}
\end{equation}
where $\langle l \rangle$ is the mean chord length and $\langle l^2 \rangle$ is the mean squared chord length. ${\langle l^2 \rangle}/{2 {\langle l \rangle}^2}$ expresses the correction for a non-exponential length distribution and  is equal to 1 for an exponential distribution. $\beta$ is a constant related to the effect of redirecting collisions between walls and oxygen molecules~\cite{henrion1977, derjaguin1946} and is given by
\begin{equation}
 \beta = 
  -{\sum^{\infty}_{m=1} \langle \cos \gamma_m \rangle},
\label{eq:twelve}
\end{equation}
where $\gamma_{m}$ is the angle between the incident and reflected directions of oxygen molecules on the wall at the $m$th collision. In the present study, the characteristic length $d$ was obtained as follows: 100\,000 lines were drawn randomly in a porous cube with a side of 200~$\upmu$m, and the length of each line segment located in the pore region was calculated as the chord length $l$. Lines were drawn in both random directions for Monte Carlo simulation and Cartesian directions to avoid counting segments smaller than the pixel size. The analysis was performed by using a center cube located at 393~$\upmu$m $\leq x \leq$ 592~$\upmu$m, 408~$\upmu$m $\leq y \leq$ 607~$\upmu$m, and 400~$\upmu$m $\leq z \leq$ 599~$\upmu$m (see Fig.~\ref{fig:div}).

The effective oxygen diffusion coefficient $D_e$ was calculated from the mean square displacement (MSD) as follows:
\begin{equation}
 D_e = 
   \epsilon {\lim_{t \rightarrow \infty} \frac{\langle {| \mathbf{r}(t)-\mathbf{r}(0) |}^2 \rangle}{6t}},
\label{eq:thirteen}
\end{equation}
where $\mathbf{r}(t)$ is the position of an oxygen molecule at time $t$. It is noted that when $D_e$ in the $x$ direction is calculated, $\mathbf{r}(t)$ should be replated by $x(t)$ and 6$t$ should become 2$t$ in Eq.~(\ref{eq:thirteen}). The calculation was carried out with a random walk Monte Carlo simulation based on the mean-square displacement method~\cite{yoshimoto2017, kaneko2020}. Specifically, 10\,000 oxygen molecules were inserted into the same porous cube of  side  200~$\upmu$m. The  molecules were allowed to move in the pore region and reflect from the pore walls following the Knudsen cosine law. Mirror boundary conditions were applied on all the surfaces of the cubic cell. The velocity of oxygen molecules was set to 481.2~m/s, which is equivalent to the mean velocity at 350~K, $\sqrt{ 8k_{B}T/{\pi}m}$, where $k_{B}$ is Boltzmann's constant, $T$ is the temperature, and $m$ is the molecular mass of oxygen. The time step of the simulation was set to  $5\times10^{-10}$~s. In this calculation condition, the mean free path between molecular collisions for an oxygen molecule is much larger than the pore size because the number density of oxygen molecules in the pore region is extremely low. Therefore, Kn can effectively be assumed to be infinite and $D_0$ can be approximated as being the same as $D_K$. The value of $D_K$ was calculated from Eqs.~(\ref{eq:five}), (\ref{eq:eleven}), and (\ref{eq:twelve}). The tortuosity factor $\tau$ was then obtained from Eqs.~(\ref{eq:two}) and (\ref{eq:thirteen}). 

\subsection{\label{sec:level23c}CFD simulation}

To calculate  air permeability through the porous sample, a CFD simulation was performed  using the X-ray CT images of porous cubes of sides  200 and 100~$\upmu$m, respectively. The air flow was assumed to be compressible, single-phase, and laminar. The conservation equations for mass and momentum are
\begin{align}
 \nabla \cdot (\rho \mathbf{u})& = 
  0,
\label{eq:fourteen}
\\[3pt]
 \rho ( \mathbf{u} \cdot \nabla ) \mathbf{u}& = 
  -\nabla P + \mu \nabla^2\mathbf{u},
\label{eq:fifteen}
\end{align}
where $\rho$ is the density of air, ${\bf u}$ is the velocity vector of air, $\mu$ is the viscosity of air, and $P$ is the pressure of air. $\rho$ was calculated by using the ideal gas law and the temperature was assumed to be 298 K.

\begin{figure}[!t]
\includegraphics[width=\columnwidth]{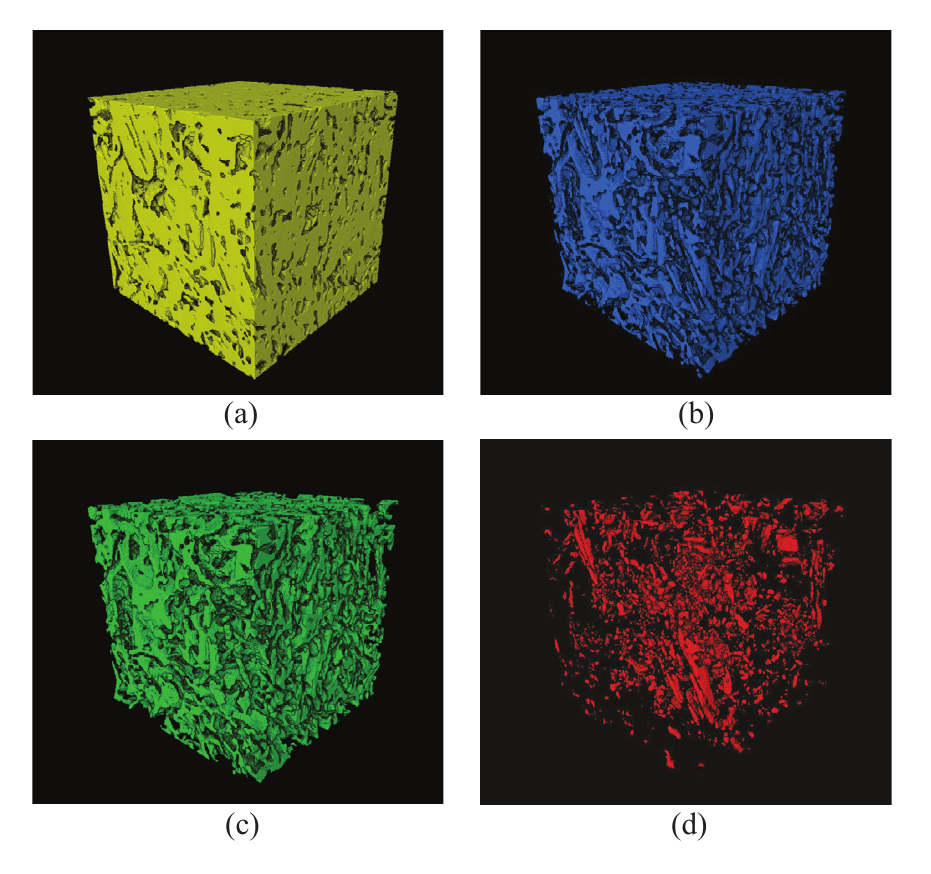}%
\caption{\label{fig:pore2} Images of the cubic analysis region with the side of 200~$\upmu$m: (a) solid; (b) void; (c)  open pore; (d) closed pores (sample A).}
\end{figure}

\begin{figure}[!t]
\includegraphics[width=\columnwidth]{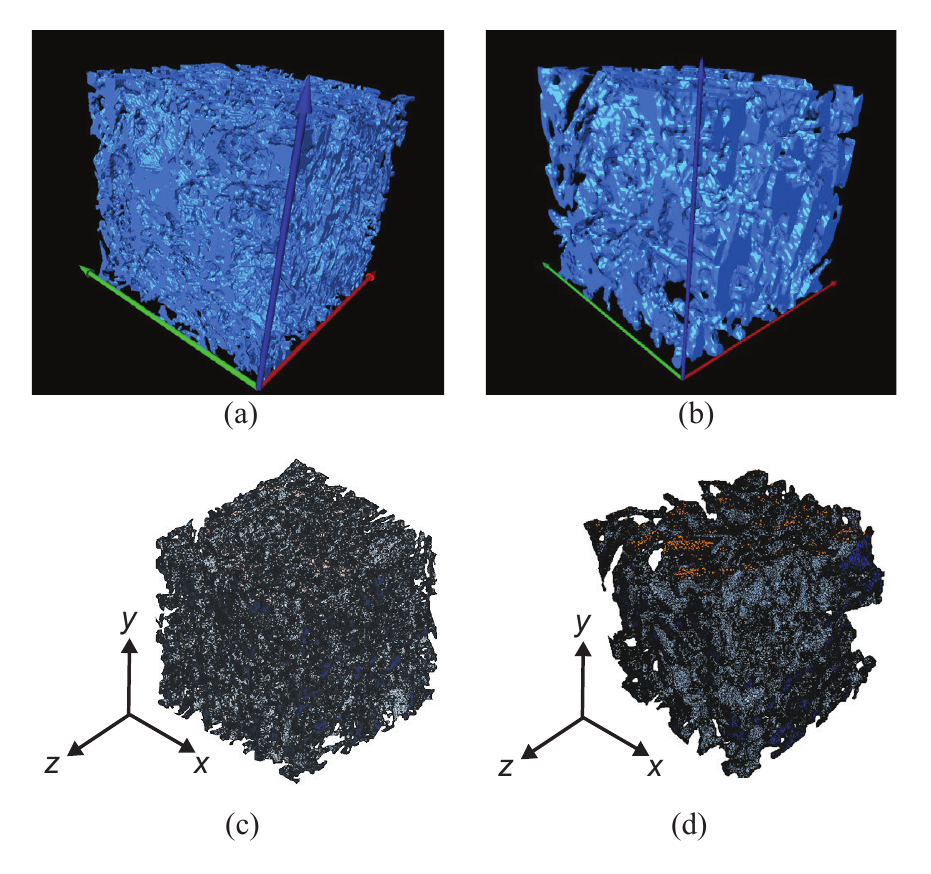}%
\caption{\label{fig:stl} STL data for  cubic analysis regions with sides of (a) 200~$\upmu$m and (b) 100~$\upmu$m, and volume meshes for cubic analysis regions with sides of (c) 200~$\upmu$m and (d) 100~$\upmu$m (sample A).}
\end{figure}

The volume mesh was created with the CT analyzer software Avizo. The center cube, located at $393~\upmu\mathrm{m} \leq x \leq 592~\upmu \mathrm{m}$, $408~\upmu \mathrm{m} \leq y \leq 607~\upmu \mathrm{m}$, and $400~\upmu \mathrm{m}\leq z \leq 599~\upmu \mathrm{m}$, was extracted from the cylindrical scanned area as shown in Fig.~\ref{fig:div}. Figure~\ref{fig:pore2} shows  images of solid, void, open pore, and closed pores in the analysis region,  corresponding  to the images for the entire scanned region shown in Fig.~\ref{fig:pore}. The volumes of solid, void, open pore, and closed pores are $5.66\times10^{6}$, $2.21\times10^{6}$, $2.13\times10^{6}$, and $8.10\times10^{4}~\upmu$m$^{3}$, respectively. From the open pore data shown in Fig.~\ref{fig:pore2}(c), stereolithographic (STL) data were created as shown in Fig.~\ref{fig:stl}(a).  STL data were also created for a smaller cube of side  100~$\upmu$m as shown in Fig.~\ref{fig:stl}(b). From these two sets of  STL data,  volume meshes with $33 \times 10^6$ and $7.3  \times 10^6$  cells were created as shown in Figs.~\ref{fig:stl}(c) and \ref{fig:stl}(d), respectively. These volume meshes had seven interfaces: the six surfaces of the cube  and the interface between  void and  solid inside the cube. With the air flow  assumed to be in the $x$ direction, the inlet and outlet of the flow were  on the two $y$--$z$ planes of the cube, and  wall boundary conditions were applied to the other four surfaces of the cube. To calculate the flow rate at a fixed pressure drop, different constant pressures were applied on the inlet and outlet boundaries. The CFD simulations were performed for three flow directions ($x$, $y$, and $z$) with a Xeon Silver 4108 CPU and Star CCM+ v14.02.010-R8 software.

%\begin{figure}[!t]
%\includegraphics[width=\columnwidth]{fig_11}%
%\caption{\label{fig:vmesh} Volume meshes for cubic analysis regions with  sides of (a) 200~$\upmu$m and (b) 100~$\upmu$m.}
%\end{figure}

\section{\label{sec:level14}RESULTS AND DISCUSSION}

\subsection{\label{sec:level24a}Measured air permeability, porosity and pore size distributions}

\begin{figure}[!t]
\includegraphics[width=\columnwidth]{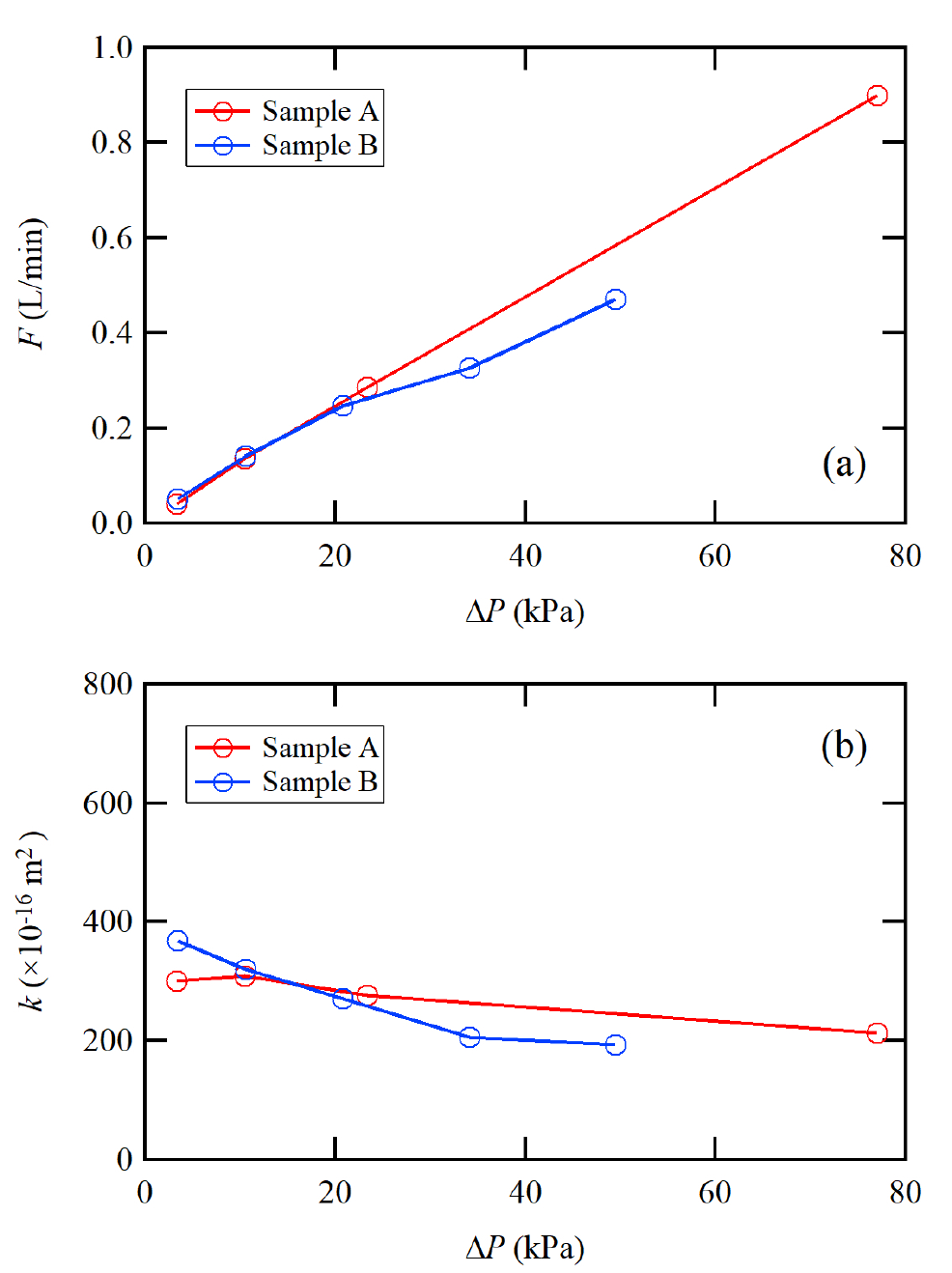}%
\caption{\label{fig:graph1} (a) Air flow rate and (b) air permeability of samples A and B measured by a capillary flow porometer.}
\end{figure}

\begin{figure}[!t]
\includegraphics[width=\columnwidth]{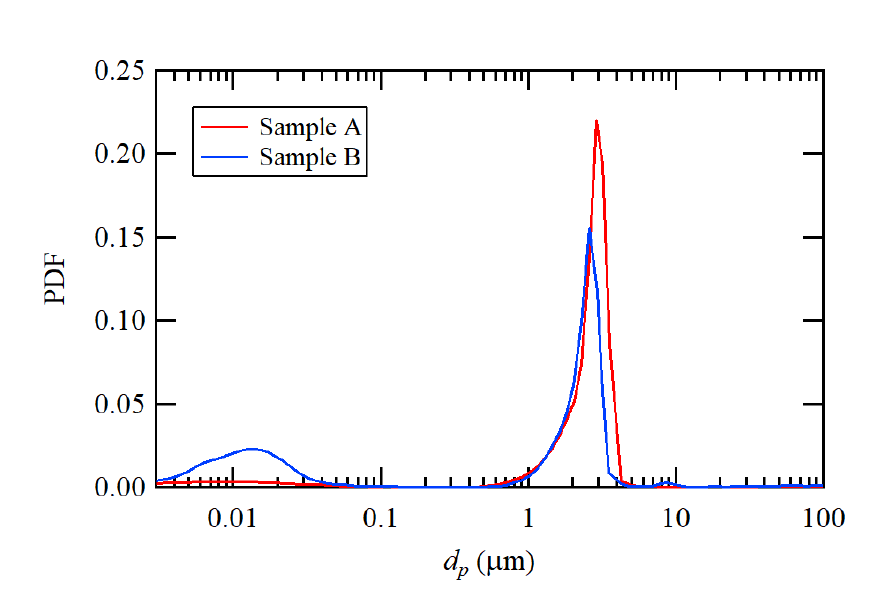}%
\caption{\label{fig:graph2} Pore size distributions of samples A and B measured by mercury porosimetry: probability density functions (PDFs) of pore diameter $d_{p}$.}
\end{figure}

The air flow rate $F$ through the porous sample measured by a capillary flow porometer is shown in Fig.~\ref{fig:graph1}(a). The measured value of $F$ increases proportionally to the pressure drop $\Delta P$ across the  plate for both samples A and B. The permeability $k$ calculated using Eq.~(\ref{eq:one}) is shown in Fig.~\ref{fig:graph1}(b). The values obtained  are of the order of $10^{-14}$~m$^2$. The probability density functions (PDFs) of pore diameter $d_{p}$ for  samples A and B measured by mercury porosimetry are shown in Fig.~\ref{fig:graph2}. Sample A has a sharp peak around $d_{p} = 3.0~\upmu$m, while sample B has a peak around $d_{p} = 2.6~\upmu$m and a second, smaller, peak around $d_{p} = 0.014~\upmu$m. Sample B includes carbon nanoparticles. The second peak around $d_{p} = 0.014~\upmu$m can be attributed to the nanopores formed by carbon nanoparticles. The measured pore volume and density of sample A were 0.164~cm$^3$/g and 1.51~g/cm$^3$, respectively, and thus the porosity was $0.164 \times1.51 = 0.248$, while the measured pore volume and density of sample B were 0.138~cm$^3$/g and 1.43~g/cm$^3$, giving a porosity of $0.138\times1.43 = 0.197$. The porosities of these two samples were also calculated using Eq.~(\ref{eq:sixteen}) from the images shown in Fig.~\ref{fig:pore}. The results are summarized in Table~\ref{tab:table3}. The porosities obtained from the two different experimental approaches and image analyses have similar values, with the porosity of sample A being 5--7\% larger than that of sample B, which led to lower solid volume and higher electrical resistivity.

\begin{table}[b]
\caption{\label{tab:table3}%
Pore volumes and porosities of  samples A and B.}
\begin{ruledtabular}
\begin{tabular}{lcc}
\textrm{}&
\textrm{Sample A}&
\textrm{Sample B}\\
\cline{2-3}
&
\multicolumn{2}{c}{X-ray CT}\\
\hline
Open pore volume ($\upmu$m$^{3}$)& $1.97 \times10^{8}$ & $1.54 \times10^{8}$\\
Total volume ($\upmu$m$^{3}$) & $7.27 \times 10^{8}$ & $7.27 \times 10^{8}$\\
Porosity $\epsilon$ & 0.271 & 0.231\\[6pt]
%\hline
&
\multicolumn{2}{c}{Mercury porosimetry}\\
\hline
Porosity $\epsilon$& 0.248 & 0.197
\end{tabular}
\end{ruledtabular}
\end{table}

\subsection{\label{sec:level24b}Calculated pore size distribution and effective oxygen diffusion coefficient}

\begin{figure}[!t]
\includegraphics[width=\columnwidth]{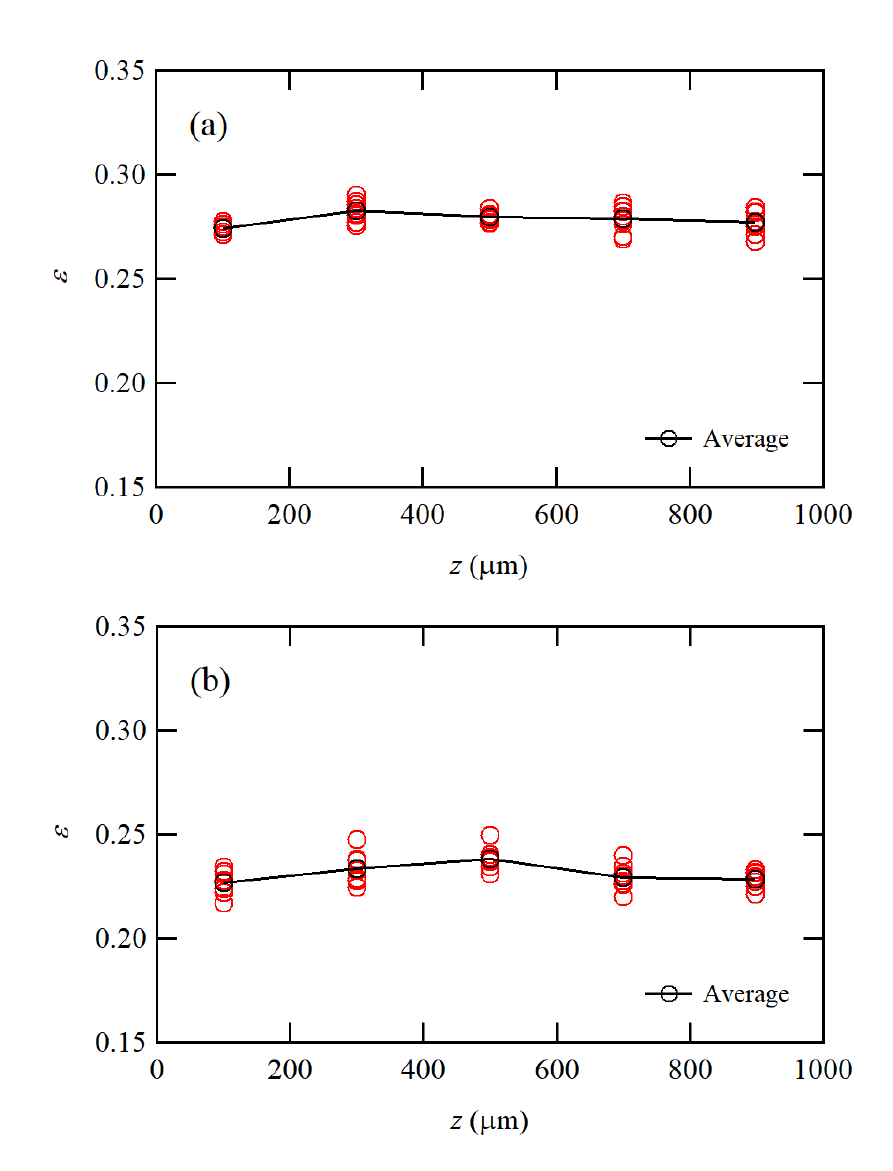}%
\caption{\label{fig:graph3} Porosity distributions of  (a) sample A and (b) sample B.}
\end{figure}

Figure~\ref{fig:graph3}(a) shows the porosity distribution of  sample A along the $z$ direction calculated from the segmented data shown in Fig.~\ref{fig:div}. From 45 segmented data ($3\times3\times 5$ cubes in the $x$, $y$, and $z$ directions, respectively), the maximum and minimum porosities are 0.290 and 0.267 respectively, which correspond to $+4.2\%$ and $-3.9\%$ of the average porosity 0.278. The porosity distribution of  sample B along the $z$ direction was also calculated and is shown in Fig.~\ref{fig:graph3}(b). These results suggest that the porosity is almost uniform in the entire scanned region.

\begin{figure}[!t]
\includegraphics[width=\columnwidth]{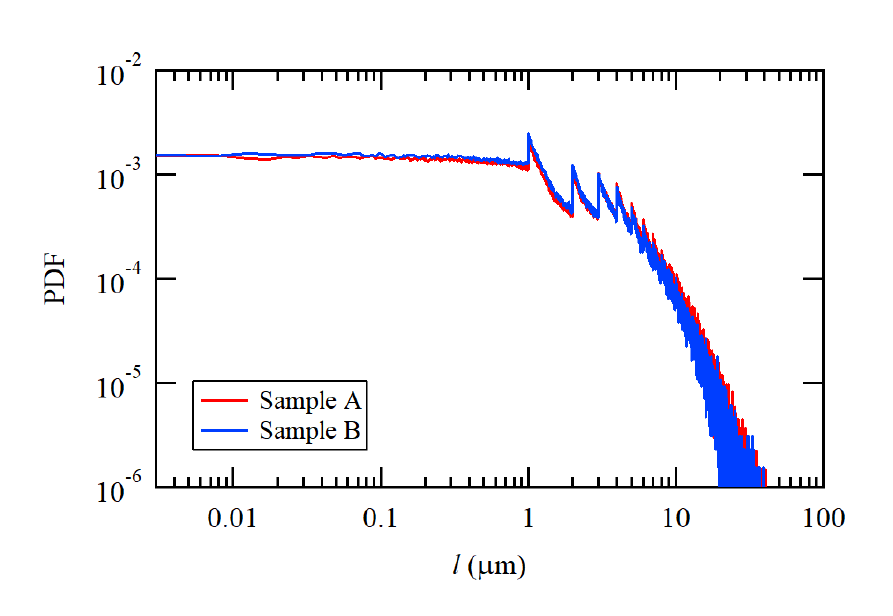}%
\caption{\label{fig:graph4} Chord length distributions in random directions of samples A and  B.}
\end{figure}

\begin{figure}[!t]
\includegraphics[width=\columnwidth]{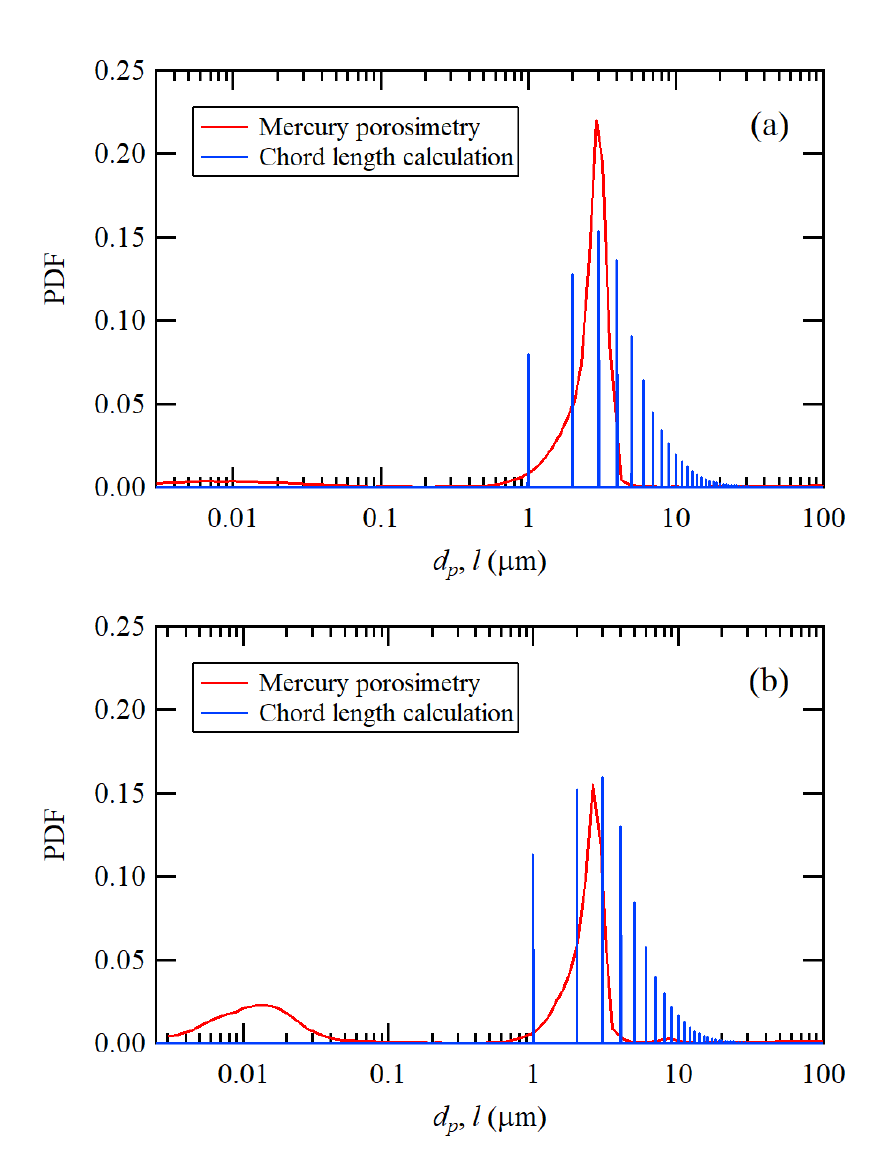}%
\caption{\label{fig:graph5} Measured pore size distributions and calculated chord length distributions in Cartesian directions of (a) sample A and  (b) sample B.}
\end{figure}

Figure~\ref{fig:graph4} shows the distribution of the chord length $l$ in random directions of the two porous samples calculated using the method described in Sec.~\ref{sec:level23b}.
Because  discrete pixel data were used for the pore structure data, a spike appears  every 1~$\upmu$m. Figure~\ref{fig:graph5} shows a comparison of the chord length distribution in Cartesian directions and the pore size distribution measured by mercury porosimetry. Although the chord length has discrete values, the peak values of the chord length and the measured pore diameter are around 3~$\upmu$m. The calculated results for the chord lengths in random directions and other variables of the porous samples are summarized in Table~\ref{tab:table4}.

\begin{table}[!t]
\caption{\label{tab:table4}%
Chord lengths in random directions and other variables of  samples A and B.}
\begin{ruledtabular}
\begin{tabular}{lcc}
&
Sample A&
Sample B\\
\hline
$d$ ($\upmu$m) & 2.46  & 2.25 \\
${\langle l \rangle}$ ($\upmu$m)& 3.65  & 3.26 \\
${\langle l^2 \rangle}$ ($\upmu$m$^{2}$) & 27.9 & 22.6 \\
${\langle l^2 \rangle}/{2 {\langle l \rangle}^2}$ & 1.05 & 1.06\\
$\beta$ & 0.371 & 0.375
\end{tabular}
\end{ruledtabular}
\end{table}

\begin{figure}[!t]
\includegraphics[width=\columnwidth]{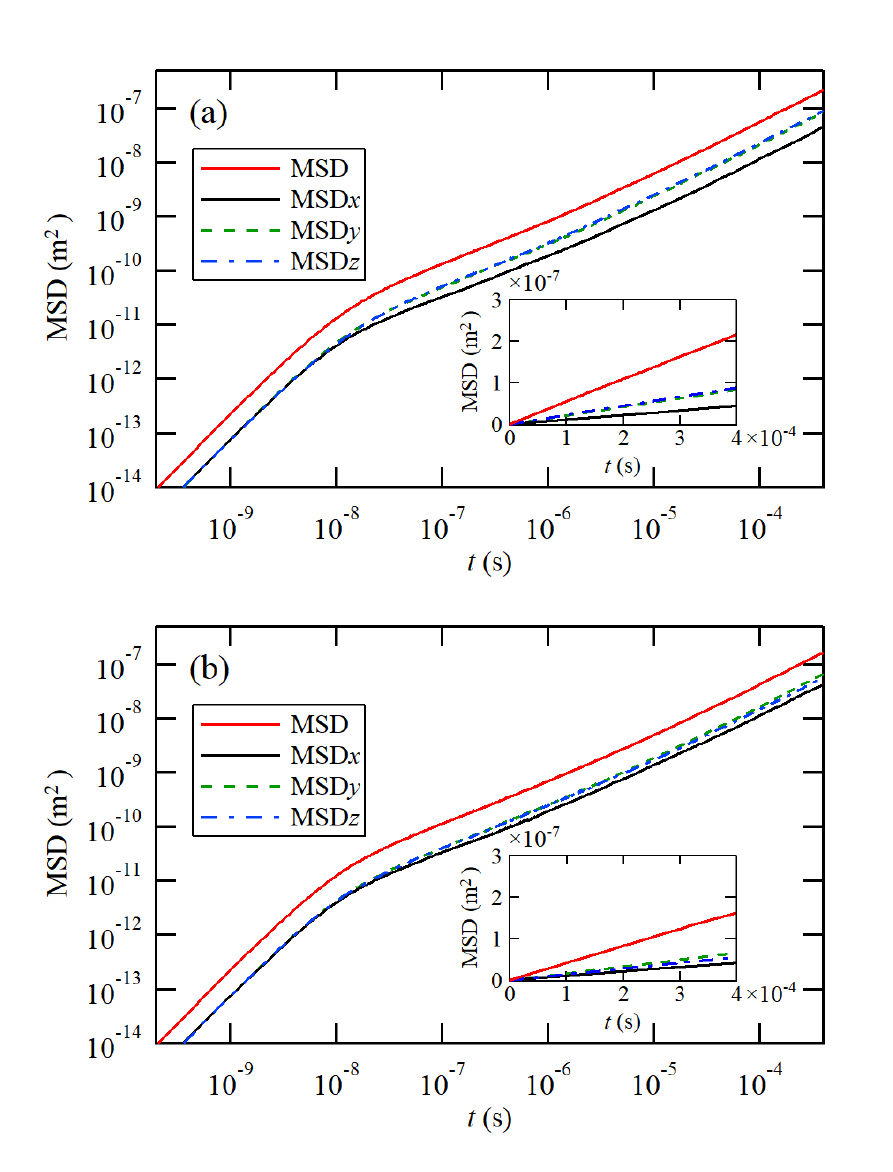}%
\caption{\label{fig:graph6} Mean square displacements (MSDs), and their $x$, $y$, and $z$ components, of oxygen molecules in (a) sample A and (b) sample B. The insets show the same graphs on a linear scale.}
\end{figure}

Figure~\ref{fig:graph6} shows the MSDs of 10\,000 oxygen molecules in the porous structure calculated by Monte Carlo simulation. The insets shows the same graphs on a linear scale. The MSD increases steeply at $t < 10^{-8}$~s because an oxygen molecule moves ballistically without colliding with the pore wall in this regime. The gradient of the MSD with respect to $t$ changes at around $t = 10^{-8}$~s. The MSD  at $t = 10^{-8}$~s is about $10^{-11}$~m$^2$, which is of the same order of magnitude as the mean squared chord length $\langle l^2 \rangle$. When $t$ is larger than $10^{-4}$~s, the MSD is proportional to $t$, and the gradient gives the value of $D_e$/$\epsilon$ [see Eq.~(\ref{eq:thirteen})]. Figure~\ref{fig:graph6} also shows the $x$, $y$, and $z$ components of the MSDs. At $t > 10^{-4}$~s, in the case of sample A, the $x$ component of the MSD is clearly lower than the other two components  [Fig.~\ref{fig:graph6}(a)], whereas for sample B, the $x$ component  is slightly less than the other two components [Fig.~\ref{fig:graph6}(b)]. This indicates that the effective diffusion coefficient of oxygen in the $x$ direction (i.e., the through-plane direction) is lower than that in the $y$ and $z$ directions (i.e., the in-plane directions). In this calculation, the number density of oxygen molecules is extremely low, and the mean free path between molecular collisions for an oxygen molecule is much larger than the pore size. Therefore, it is expected that the effective diffusion coefficient of oxygen largely depends on the porous structure, and the tortuosity factor, which represents the tortuous nature of a porous medium, can be evaluated appropriately. The calculated tortuosity factors $\tau$ are 4.4 and 5.8 for samples A and B, respectively. These values are slightly higher than the reference values of a porous material with a similar porosity of around 0.3~\cite{kinefuchi2013}. The calculated diffusion coefficients, porosities, and tortuosity factors are summarized in Table~\ref{tab:table5}. 

\begin{table}[!t]
\caption{\label{tab:table5}%
Oxygen diffusion coefficients and porosity and tortuosity factors of  samples A and B.}
\begin{ruledtabular}
\begin{tabular}{lcc}
&
Sample A&
Sample B\\
\hline
$D_e$ (m$^{2}$/s) & $2.52 \times10^{-5}$  & $1.64 \times10^{-5}$\\
$D_e(x~\mathrm{direction})$ (m$^{2}$/s) & $1.56 \times 10^{-5}$  & $1.26 \times10^{-5}$  \\
$D_e(y~\mathrm{direction})$ (m$^{2}$/s) & $2.93 \times10^{-5}$  & $1.99 \times10^{-5}$  \\
$D_e(z~\mathrm{direction})$ (m$^{2}$/s) & $3.08 \times10^{-5}$  & $1.67 \times10^{-5}$  \\
$D_K$ (m$^{2}$/s) & $3.95 \times10^{-4}$  & $3.60 \times10^{-4}$ \\
$\epsilon$ & 0.281 & 0.240\\
$\tau$ & 4.40 & 5.28
\end{tabular}
\end{ruledtabular}
\end{table}

\subsection{\label{sec:level24c}Calculated air permeability}

\begin{figure}[!t]
\includegraphics[width=\columnwidth]{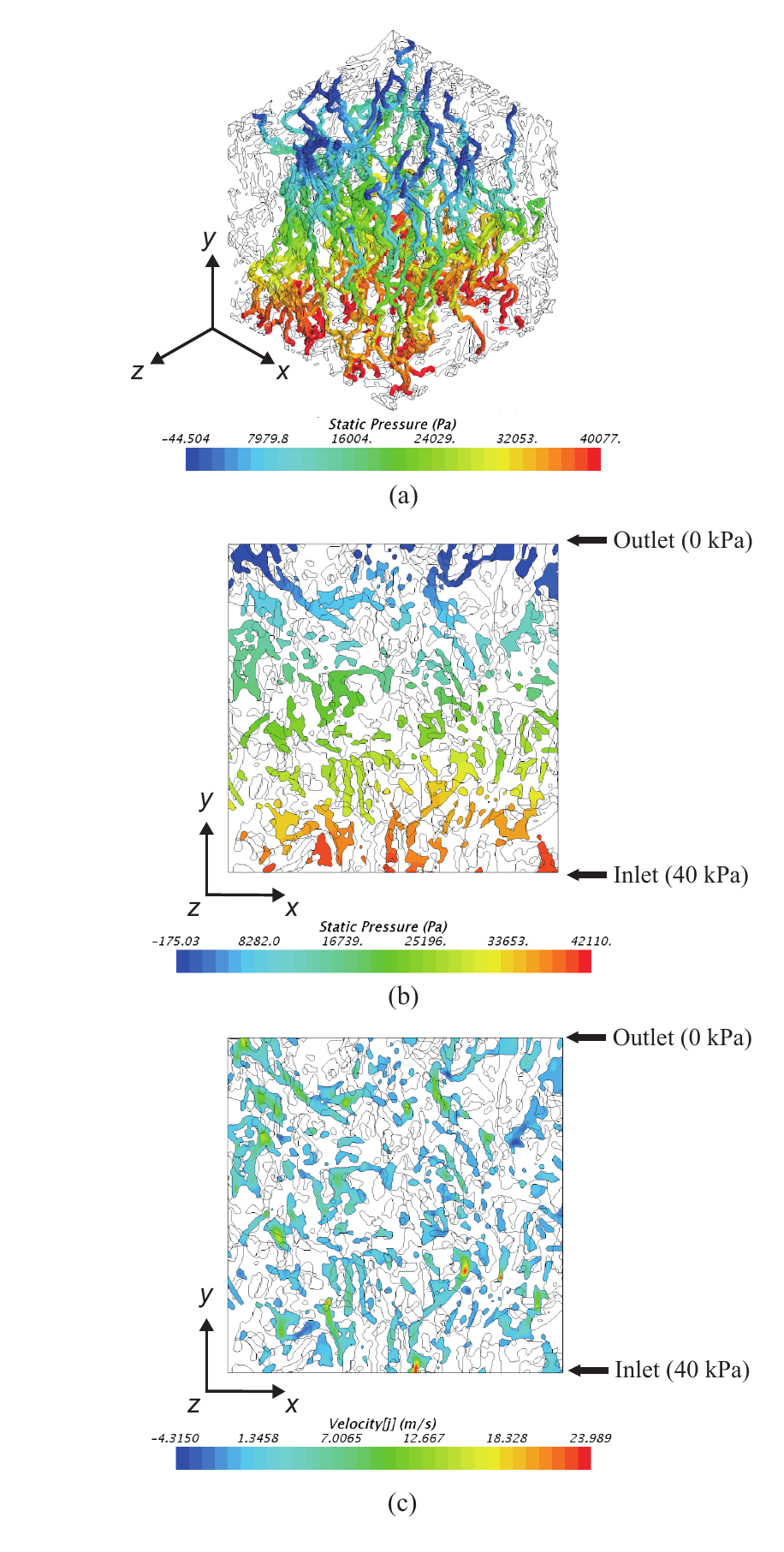}%
\caption{\label{fig:cfd} (a) Streamlines, (b) pressure distribution, and (c) velocity magnitude distribution for the flow in the $y$ direction at $\Delta P = 40$~kPa (sample A).}
\end{figure}

CFD simulations of  air flow through sample A were performed with the volume mesh shown in Fig.~\ref{fig:stl}(c). Figure~\ref{fig:cfd} shows streamlines and pressure and flow velocity magnitude distributions when $\Delta P = 40$~kPa was applied between two surfaces facing each other in the $y$ direction for the volume mesh of a cube of  side  200~$\upmu$m. The velocity magnitude distribution appears to be uniform, but the average velocity slightly increaces from the inlet to the outlet. The velocities at the inlet and the outlet are 1.8 m/s and 2.2 m/s, respectively.

The volumetric flow rate and permeability [Eq.~(\ref{eq:one})] of sample A were calculated when $\Delta P = 40 $~kPa was applied in the $x$, $y$, and $z$ directions, respectively. The  results are summarized in Table~\ref{tab:table6}. The volumetric flow rate and permeability of sample A in the $x$ direction are about half of those in the $y$ and $z$ directions. In Sec.~\ref{sec:level24b}, it was shown that the oxygen diffusivity of sample A in the $x$ direction was about  half of that in the $y$ and $z$ directions. These anisotropic transport properties can be attributed to the anisotropic porous structure of sample A. Because the flat porous separator is made by pouring the mixture of carbon particles and resin into a mold and pressing in the $x$ direction, the porous structure should be compressed in the $x$ direction, and thus the transport in that direction should be suppressed. It is fair to say that the anisotropic transport properties have been successfully reproduced by the Monte Carlo calculation of oxygen diffusivity and the CFD calculation of air permeability based on the X-ray CT images of the porous structure.

\begin{table}[!t]
\caption{\label{tab:table6}%
Volumetric flow rates and permeability of air in sample A obtained with cubic volume meshes of  sides  100 and 200~$\upmu$m at $\Delta P = 40$~kPa.}
\begin{ruledtabular}
\begin{tabular}{lccc}
&
\multicolumn{3}{c}{$100\times100\times100~\upmu$m$^{3}$}\\
\cline{2-4}
&
\multicolumn{2}{c}{Volumetric flow rates }
&\\
&
Inlet (m$^{3}$/s) & Outlet (m$^{3}$/s) & Permeability (m$^{2}$)\\
\hline
$x$ direction  & $3.69  \times 10^{-9}$
& $4.90  \times 10^{-9}$ & $1.87  \times 10^{-14}$ \\
$y$ direction  & $6.35  \times 10^{-9}$
& $8.26  \times 10^{-9}$ & $3.14  \times 10^{-14}$ \\
$z$ direction  & $8.77  \times 10^{-9}$
& $1.17  \times 10^{-8}$ & $4.46  \times 10^{-14}$  \\[6pt]
&
\multicolumn{3}{c}{$200\times200\times200~\upmu$m$^{3}$}\\
\cline{2-4}
&
\multicolumn{2}{c}{Volumetric flow rates }
&\\
&
Inlet (m$^{3}$/s) & Outlet (m$^{3}$/s) & Permeability (m$^{2}$)\\
\hline
$x$ direction  & $8.88  \times 10^{-9}$
& $1.17  \times 10^{-8}$ & 2.24  $\times 10^{-14}$  \\
$y$ direction  & $1.67  \times 10^{-8}$
& $2.21  \times 10^{-8}$ & 4.19  $\times 10^{-14}$  \\
$z$ direction  & $1.81  \times 10^{-8}$
& $2.44  \times 10^{-8}$ & 4.65  $\times 10^{-14}$ 
\end{tabular}
\end{ruledtabular}
\end{table}

\begin{table}[t]
\caption{\label{tab:table7}%
Volumetric flow rates and permeability of air in samples A and B obtained with a cubic volume mesh of side  100~$\upmu$m at $\Delta P = 20$ and 40~kPa.}
\begin{ruledtabular}
\begin{tabular}{lccc}
&
\multicolumn{3}{c}{Sample A at $\Delta P = 20$~kPa}\\
\cline{2-4}
&
\multicolumn{2}{c}{Volumetric flow rates }
&\\
&
Inlet (m$^{3}$/s) & Outlet (m$^{3}$/s) & Permeability (m$^{2}$)\\
\hline
$x$ direction & $2.01  \times 10^{-9}$
& $2.28  \times 10^{-9}$ & $1.89  \times 10^{-14}$  \\
$y$ direction & $3.50  \times 10^{-9}$
& $3.87  \times 10^{-9}$ & $3.21  \times 10^{-14}$  \\
$z$ direction & $4.81  \times 10^{-9}$
& $5.49  \times 10^{-9}$ & $4.55  \times 10^{-14}$  \\[6pt]
&
\multicolumn{3}{c}{Sample B at $\Delta P = 20$~kPa}\\
\cline{2-4}
&
\multicolumn{2}{c}{Volumetric flow rates }
&\\
&
Inlet (m$^{3}$/s) & Outlet (m$^{3}$/s) & Permeability (m$^{2}$)\\
\hline
$x$ direction & $2.26  \times 10^{-9}$
& $2.50  \times 10^{-9}$ & $2.07  \times 10^{-14}$  \\
$y$ direction & $3.20  \times 10^{-9}$
& $3.60  \times 10^{-9}$ & $2.98  \times 10^{-14}$  \\
$z$ direction & $2.38  \times 10^{-9}$
& $2.71  \times 10^{-9}$ & $2.24  \times 10^{-14}$  \\[6pt]
&
\multicolumn{3}{c}{\textrm{Sample B at $\Delta P$ = 40 kPa}}\\
\cline{2-4}
&
\multicolumn{2}{c}{Volumetric flow rates }
&\\
&
Inlet (m$^{3}$/s) & Outlet (m$^{3}$/s) & Permeability (m$^{2}$)\\
\hline
$x$ direction & $4.13  \times 10^{-9}$
& $5.27  \times 10^{-9}$ & $2.00  \times 10^{-14}$  \\
$y$ direction & $5.84  \times 10^{-9}$
& $7.71  \times 10^{-9}$ & $2.93  \times 10^{-14}$  \\
$z$ direction & $4.36  \times 10^{-9}$
& $5.81  \times 10^{-9}$ & $2.21  \times 10^{-14}$ 
\end{tabular}
\end{ruledtabular}
\end{table}

\begin{figure}[!t]
\includegraphics[width=\columnwidth]{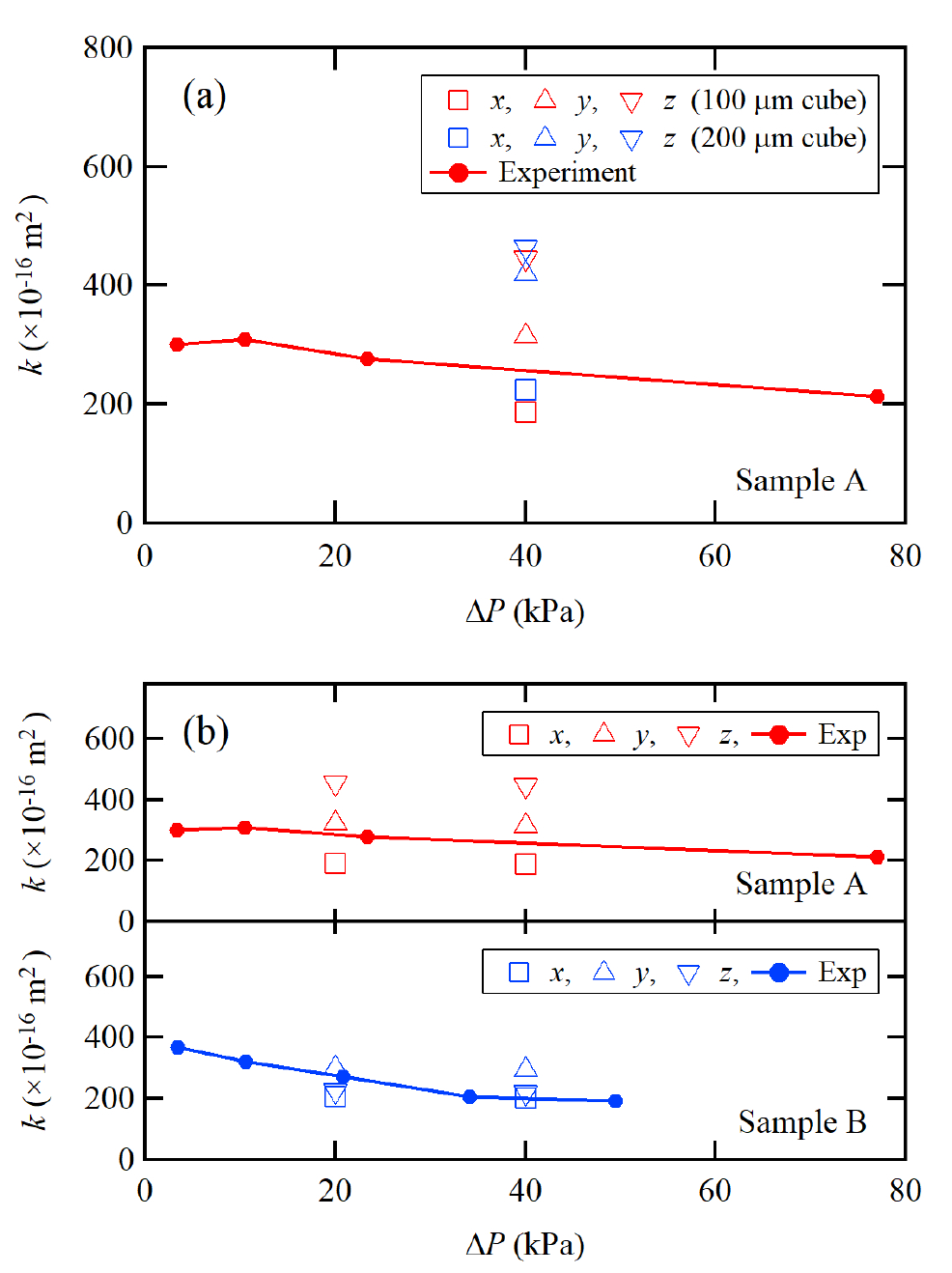}%
\caption{\label{fig:graph7} Comparison between calculated and experimental results for air permeabilities in the $x$, $y$, and $z$ directions: (a) effect of analysis region size; (b) effect of sample type for a porous cube of side 100~$\upmu$m.}
\end{figure}

Similar CFD simulations of  air flow through sample A at $\Delta P = 40$~kPa were performed with the cubic volume mesh of side  100~$\upmu$m shown in Fig.~\ref{fig:stl}(d). The calculated results for the volumetric flow rate and permeability of air in the $x$, $y$, and $z$ directions are shown in Table~\ref{tab:table6}. Figure~\ref{fig:graph7}(a) shows a comparison between the calculated air permeability for the two  sizes of analysis region and the measured air permeability. All the values of air permeability are of the order of $10^{-14}$~m$^{2}$. The calculated results for both sizes of analysis region show a similar trend, namely, that the volumetric flow rate and permeability in the $x$ direction are about  half of those in the $y$  and $z$ directions.

Further simulations were performed using a cubic volume mesh of side  100~$\upmu$m. The same CFD simulations of  air flow through a porous sample were performed at $\Delta P = 20$~kPa for sample A and at $\Delta P = 20$ and 40~kPa for sample B. The calculated results for the volumetric flow rate and permeability of air in the $x$, $y$ and $z$ directions are shown in Table~\ref{tab:table7}. In comparison between $\Delta P = 20$ and 40~kPa for each sample, the volumetric flow rate at $\Delta P = 40$~kPa is about twice larger than that at $\Delta P = 20$~kPa and the permeability in the $x$, $y$ and $z$ directions are almost the same both at $\Delta P = 20$ and 40~kPa. Figure~\ref{fig:graph7}(b) shows a comparison of permeability between samples A and B, and it can be seen that the calculated permeability agrees with the experimental one. The permeability in the $x$ direction of sample B is close to that of sample A, while the porosity of sample B is lower than that of sample A, as shown in Table~\ref{tab:table3}. This suggests that the anisotropy of the porous structure in sample B is not as great as that in sample A. As already mentioned, sample B is a low electrical resistivity type. Because sample B includes carbon nanoparticles, the pore size distribution has a second, smaller, peak around $d_{p} = 0.014~\upmu$m (see Fig.~\ref{fig:graph2}). The carbon nanoparticles can affect the pore formation process during the solidification of carbon particles with a binder resin in a mold. The mechanism leading to the formation of an anisotropic pore structure is still to be determined, but the air permeability in the $y$ and $z$ directions was suppressed to the same degree as in the $x$ direction in sample B. The proposed  method for evaluating gas permeability using CFD simulation based on micro X-ray CT images makes it possible to quantify gas permeability in a given direction inside an anisotropic porous medium.

\section{\label{sec:level15}CONCLUSION}

A method to evaluate gas transport properties in porous structures has been proposed. This method involves the following procedures: (1)  observation of local pore structures with X-ray CT; (2)  creation of a 3D porous structure from the CT images; (3)  analysis of gas diffusion coefficients and gas permeability in   porous structures using random walk Monte Carlo simulations and CFD simulations, respectively. To validate the calculated pore size distribution and gas permeability, these values were also experimentally measured by mercury porosimetry and a capillary flow porometer, respectively. Specifically,  3D X-ray CT was used to observe the internal structure of porous   PEFC separators of different electrical resistivities with a spatial resolution of 1~$\upmu$m per pixel over a cylindrical region of diameter 1~mm and height 1 mm. The pore structural data were  extracted from binarized image files, and the porosity was compared with the measured value. To evaluate the porous structure, the chord length distribution was calculated and compared with the pore size distribution obtained from mercury porosimetry. The effective oxygen diffusion coefficient in the pore structure was also calculated by random walk Monte Carlo simulations, and the difference between the in-plane and through-plane components was discussed. Finally, CFD simulations of  air flow through porous samples were performed, and the volumetric flow rate and permeability of air were compared with those obtained from flow porometry. 

The following conclusions can be drawn from this study.

The porosity calculated from the X-ray CT images [Eq.~(\ref{eq:sixteen})] was in good agreement with that measured by mercury porosimetry. A 600~$\upmu$m $\times$ 600~$\upmu$m $\times$ 1000~$\upmu$m  rectangular region was divided into 45 cubic regions each of 200~$\upmu$m $\times$ 200~$\upmu$m $\times$ 200~$\upmu$m, and the porosity distribution was evaluated. The porosity of each cubic region was within $\pm$4.2\% of the average value. Although the chord length had discrete values, its peak value was in  good agreement with the peak pore diameter of around 3~$\upmu$m measured by mercury porosimetry.

The MSD of oxygen in the $x$ direction (i.e., the through-plane direction) was clearly lower than those in the $y$ and $z$ directions (i.e., the in-plane directions) for sample A (the sample with conventional resistivity). The volumetric flow rate and permeability of air in the $x$ direction were also clearly lower than those in the $y$ and $z$ directions for sample A. These anisotropic transport properties can be attributed to the anisotropic porous structure of this sample. Because the flat porous separator was pressed in the $x$ direction during the solidification process of carbon particles with a binder resin in a mold, the porous structure should be compressed in that direction. Thus, anisotropic transport properties were successfully reproduced by  Monte Carlo calculation of diffusivity and CFD calculation of permeability based on  X-ray CT images of the porous structure.

The MSD of oxygen in the $x$ direction (i.e., the through-plane direction) was slightly lower than those in the $y$ and $z$ directions (i.e., the in-plane directions) for sample B (the low-resistivity sample). The air permeability in the $x$ direction of sample B was close to that of sample A, while the porosity of sample B was lower than that of sample A. These results suggest that sample B was solidified more isotropically and more tightly. As a result, the air permeabilities in the $x$, $y$, and $z$ directions were suppressed similarly. Thus proposed  method using CFD simulation based on micro X-ray CT images makes it possible to evaluate anisotropic gas permeabilities in anisotropic porous media.

\bibliography{apssamp_rev}% Produces the bibliography via BibTeX.

\end{document}